\documentclass[aps,nofootinbib,showpacs,showkeys,final,balancelastpage,superscriptaddress]{revtex4}
\usepackage{amssymb}
\usepackage{amsmath}
\usepackage{amsbsy}
\usepackage{amsfonts}
\usepackage{graphicx}

\DeclareFontFamily{OT1}{pzc}{}
\DeclareFontShape{OT1}{pzc}{m}{it}{<-> s * [1.10] pzcmi7t}{}
\DeclareMathAlphabet{\mathpzc}{OT1}{pzc}{m}{it}
\providecommand{\U}[1]{\protect\rule{.1in}{.1in}}
\DeclareFontFamily{OT1}{pzc}{}
\DeclareFontShape{OT1}{pzc}{m}{it}{<-> s * [1.10] pzcmi7t}{}
\DeclareMathAlphabet{\mathpzc}{OT1}{pzc}{m}{it}
\providecommand{\U}[1]{\protect\rule{.1in}{.1in}}
\providecommand{\U}[1]{\protect\rule{.1in}{.1in}}
\providecommand{\U}[1]{\protect\rule{.1in}{.1in}}
\providecommand{\U}[1]{\protect\rule{.1in}{.1in}}
\providecommand{\U}[1]{\protect\rule{.1in}{.1in}}
\input{tcilatex}
\begin{document}

\title{The derivation of model kinetic equation for gases and for plasmas}
\author{ Viacheslav \surname{V. Belyi}}
\affiliation{IZMIRAN, Russian Academy of Siences, Troitsk, Moscow, Russia}
\email{s b e l y i @ i z m i r a n . r u}

\begin{abstract}
A new form of the model collision operator for a Boltzmann gas of hard
spheres and Coulomb plasma is derived. One-component and many-component
systems are considered. The collision operator proposed takes properly into
account the relaxation of the first 13 hydrodynamic moments. An expression
for the intensity of the Langevin source in the model kinetic equation is
obtained in the same approximation. A technique for reconstruction of the
model collision integral based on a known expression for the model
linearized operator is proposed. It is shown that, within our model, the
collision integral does not contain a complicated exponential, common for
the ellipsoidal statistical type models. Boltzmann's H-theorem is proved for
our model.
\end{abstract}

\maketitle


\section{Introduction}

As it is well known, neither the Boltzmann kinetic equation for gases nor
the Landau or Balescu-Lenard equations for a plasma can be resolved exactly
and one uses approximations and models that preserve the essential
properties of the original collision operator. The most widely used model
kinetic equation, especially efficient in the case of discrete simulation,
for example in the lattice Boltzmann gas calculations \cite{Karlin}, is the
Bhatnagar, Gross and Krook (BGK) model \cite{BGK}. We recall that in the BGK
model the collision term for the one-component system%
\begin{equation}
I^{BGK}\{f\}=-\nu (f-f^{0})  \label{a.6}
\end{equation}%
is the deviation of the distribution function (d.f.) $f\ $from the
Maxwellian d.f. 
\begin{equation}
f^{0}=\frac{n}{(2\pi mT)^{3/2}}\exp -\frac{m(\mathbf{v-V})^{2}}{2T}
\label{a.7}
\end{equation}%
whose parameters: $n(\mathbf{r},t)=\int fd\mathbf{p;}$ $\mathbf{V}(\mathbf{r}%
,t)=\frac{1}{n}\int \mathbf{v}fd\mathbf{p;}$ $\mathbf{T}(\mathbf{r},t)=\int 
\frac{m(\mathbf{v-V})^{2}}{3n}fd\mathbf{p}$ - the local density, the mean
velocity and the temperature in energy units, are, respectively, moments of
the distribution function $f$. The term (\ref{a.6}) vanishes at equilibrium
and satisfies the conservation laws:

\begin{equation}
\int \varphi (\mathbf{p})I^{BGK}\{f\}d\mathbf{p}=0,\text{ if }\varphi (%
\mathbf{p})=1,\text{ }\mathbf{p},\text{ }\frac{\mathbf{p}^{2}}{2m}
\label{a.9}
\end{equation}%
and Boltzmann's H-theorem \cite{Cerc}:%
\begin{equation}
\frac{\partial }{\partial t}H^{BGK}(t)=\nu \int (f-f^{0})\log \frac{f}{f^{0}}%
d\mathbf{p}\leq 0.  \label{a.10}
\end{equation}%
In the problems of linear transport and fluctuations, one commonly uses the
linearized form of the BGK collision operator:%
\begin{equation}
{\delta }\widehat{I}\left\vert h\right\rangle {\ =-\nu }\left( \left\vert
h\right\rangle {\ -}\sum_{\alpha =1}^{5}\left\vert \Psi _{\alpha
}\right\rangle \left\langle \Psi _{\alpha }|h\right\rangle \right) ,
\label{a.11}
\end{equation}%
where $\left\vert h\right\rangle $ is defined by ${\ f=f}^{0}{\ +\delta f=f}%
^{0}{\ (1+h),}$ and $\left\vert \Psi _{\alpha }\right\rangle $ are the first
five Hermite polynomials.

The advantage of the BGK model is that the solution of the kinetic equation
reduces to that of a system of algebraic equations \cite{Cerc}. A weak point
is that the model implies that the Prandtl number ($\Pr $) equals 1, while
for monatomic gases the Prandtl number\ amounts to 2/3.

Holway \cite{Hol} introduced the ellipsoidal statistical model in order to
take into account real Prandtl numbers by substituting the local anisotropic
Gaussian distribution for the local Maxwellian distribution:

\begin{equation*}
f^{0}=n\pi ^{-3/2}(\det A)^{1/2}\exp -\sum_{i,j=1}^{3}\alpha _{ij}(\mathbf{v}%
_{i}\mathbf{-V}_{i})(\mathbf{v}_{j}\mathbf{-V}_{j})\mathbf{;}
\end{equation*}%
\begin{equation}
A=\left\Vert \alpha _{ij}\right\Vert =\left\Vert (\Pr )^{-1}(2T/m)\delta
_{ij}-2(1-\Pr )P_{ij}/n\Pr \right\Vert ^{-1}.  \label{a.12}
\end{equation}%
{%
\begin{equation*}
P_{ij}=m\dint {d\mathbf{p}}f[(\mathbf{v}_{i}\mathbf{-V}_{i})(\mathbf{v}_{j}%
\mathbf{-V}_{j})-{\delta }_{ij}\frac{(\mathbf{v-V})^{2}}{3}]\text{ }
\end{equation*}%
is the pressure tensor}.

Similar problems arise in the calculation of fluctuation characteristics in
gases and plasmas. As it is well known, the fluctuations of the distribution
function can be determined by a fluctuation kinetic equation of Boltzmann
type (for neutral particles) or Landau or Balescu-Lenard type (for plasma)
with additional random source, called by analogy with Brownian motion the
"Langevin source". The intensity of this Langevin source in an equilibrium
one-component system is determined by the linearized collision operator

\begin{equation}
(yy)_{\omega ,\mathbf{k,p}_{1},\mathbf{p}_{2}}=-({\delta }\widehat{I}_{%
\mathbf{p}_{1}}+{\delta }\widehat{I}_{\mathbf{p}_{2}})f^{0}(\mathbf{p}_{1}){%
\delta (}\mathbf{p}_{1}-\mathbf{p}_{2}).  \label{a.13}
\end{equation}%
The kinetic theory of such fluctuations for Boltzmann gases was developed
for the first time by Kadomtsev \cite{Kadom}. The most complete survey of
studies for nonequilibrium electron gases was given in \cite{Kogan, GGK}.
The theory of kinetic fluctuations in nonequilibrium multicomponent gases
and plasmas was developed by Klimontovich \cite{Kli}. Such a Langevin
approach, which is widely used in different fields, is very convenient for
calculating fluctuation characteristics. For example, the spectral function
of the fluctuations of distribution function can be expressed in terms of
the intensity of the Langevin source (\ref{a.13}) and Green's function of
the linearized kinetic equation. The intensity of the Langevin source must
be found in the same approximation as the solution of the linearized kinetic
equation with collisions. In the BGK approximation, the intensity of the
Langevin source has the form:

\begin{equation}
(yy)_{\omega ,\mathbf{k,p}_{1},\mathbf{p}_{2}}=2\nu f^{0}(\mathbf{p}_{1})\{{%
\delta (}\mathbf{p}_{1}-\mathbf{p}_{2})-f^{0}(\mathbf{p}_{2})[1+\frac{m{%
\delta }\mathbf{v}_{1}{\delta }\mathbf{v}_{2}}{T}+\frac{1}{6}\left( 3-\frac{m%
{\delta }\mathbf{v}_{1}^{2}}{T}\right) \left( 3-\frac{m{\delta }\mathbf{v}%
_{2}^{2}}{T}\right) ]\};\text{ }{\delta }\mathbf{v=v-V.}  \label{a.14}
\end{equation}%
The expression (\ref{a.14}) has the invariant properties

\begin{equation}
\int \Phi (\mathbf{p}_{1})\Psi (\mathbf{p}_{2})(yy)_{\omega ,\mathbf{k,p}%
_{1},\mathbf{p}_{2}}d\mathbf{p}_{1}d\mathbf{p}_{2}=0  \label{a.15}
\end{equation}%
for $\Phi $ or $\Psi =1,\mathbf{p,p}^{2}/2m$, but it does not give correct
values for the intensities of Landau-Lifshitz Langevin sources in
hydrodynamic equations \cite{Landau}. In such a description, the intensities
of the external stress tensor and the external heat flux vector are
determined by the same relaxation frequency ${\nu }$.

A graver situation arises in the case of many-component systems. According
to the Gross and Krook (GK) model \cite{GK}, the collision operator has the
form of the deviation of d.f. from a \textquotedblleft
mythical\textquotedblright\ exponent:

\begin{equation}
I_{a}^{GK}\{f_{a}\}=-\sum_{b}\nu _{ab}[f_{a}-\frac{n_{a}}{(2\pi
m_{a}T_{ab})^{3/2}}\exp -\frac{m_{a}(\mathbf{v-V}_{ab})^{2}}{2T_{ab}}],
\label{a.16}
\end{equation}%
where the parameters $\mathbf{V}_{ab}$ and $T_{ab}$ are related linearly to
the d.f. moments $\mathbf{V}_{a};$ $\mathbf{V}_{b};$ $T_{a};$ $T_{b}:$%
\begin{equation}
\mathbf{V}_{ab}=\alpha _{aa}\mathbf{V}_{a}+\alpha _{ab}\mathbf{V}_{b},\text{ 
}T_{ab}=\beta _{aa}T_{a}+\beta _{ab}T_{b}.  \label{a.17a}
\end{equation}%
Coefficients $\alpha _{aa}\mathbf{,}$ $\alpha _{ab},$ $\beta _{aa}$ and $%
\beta _{ab}$ are chosen in such manner that both the conservation laws and
balance equations for the momenta and energy for each component hold valid.
Since the number of equations to be satisfied by the parameters of the model
(for the five-moment description of a two-component system there are four
equations: two for the balance of moments and two for the balance of
temperature) is less than the number of unknown parameters (in this
approximation these are five: $\nu _{ab}$; $\alpha _{aa}\mathbf{;}$ $\alpha
_{ab};$ $\beta _{aa}$; $\beta _{ab}$), there is an arbitrariness in the
choice of parameters. Therefore there exist various modifications (see, for
example, \cite{Green}) of the collision model, which correctly describe
relaxation of the five moments. But, probably, the most dubious point of the
GK model is the complicated exponential dependence on d.f. Recently, a new
type of G-K, so-called ellipsoidal statistical model for gas mixtures \cite%
{Aoki, Brull, Spiga}, was proposed ad hoc and\ the Boltzmann's H- theorem
was proven for this model. However, the conservation laws and the H-theorem
are but a necessary, and not a sufficient conditions for a model to be
satisfactory. The correct model should be derived. One of the first works
dedicated to derivation of the model linearized collision integral is the
paper by Gross and Jackson (GJ) \cite{GJ}. Later, extension of the BGK
kinetic model for the inclusion of higher order matrix elements was
discussed and it was applied to investigate the generalized Enskog equation
and the dynamic structure factor for gas and fluids \cite{Furtado76,
Dufty79, Cohen87, Dufty90, Dufty98}. The approximation consisted in taking
into account exactly a finite part of the matrix operator, while the
remaining part was only represented by the diagonal matrix elements. In the
present paper we do not consider spatial inhomogeneities and assume the wave
vector k =0. But we do take into account the non-diagonal components arising
in the collision operator expansion with respect to the complete system of
polynomials in the quadratic approximation. Making use of these non-diagonal
elements allows us to obtain a new form for the model integral for Coulomb
plasma with transport coefficient correction, comparable with the Spitzer
corrections. In the case of a Boltzmann gas of hard spheres these
corrections are insignificant. An expression for the intensity of the
Langevin source in the model kinetic equation is obtained in the same
approximation. From this expression follow correct values for the
intensities of Landau-Lifshitz Langevin sources in hydrodynamic equations.
Using the technique developed for an one-component system, a consistent
derivation of the model linearized collision operator for a many-component
system is presented. In these results an ambiguity in the choice of
coefficients is eliminated, in contrast to the GK type models. A technique
for reconstruction of the form of the model collision integral based on a
known expression for the model linearized operator is proposed. It is shown
that the model collision integral in the local (not complete) equilibrium
approximation does not contain a complicated exponential, that is common for
the GK type integrals. Boltzmann's H-theorem is proved for our model.

\section{MODEL CONSTRUCTION}

\subsection{One-component systems}

To correct the BGK model following Sirovich \cite{Sirovich} we introduce two
projection operators $\widehat{H}$ and $\widehat{N}$ satisfying

{%
\begin{equation}
\widehat{H}\widehat{N}=\widehat{N}\widehat{H}=0;\ \widehat{H}+\widehat{N}=%
\widehat{Id}.  \label{b.0}
\end{equation}%
} Here $\widehat{Id}$ is the identity operator, $\widehat{H}$\ is the
operator of projection onto the 'hydrodynamic subspace' spanned by kets
corresponding to the polynomials of the lowest order in the moment variable.
In the BGK model these kets are the first five polynomials which correspond
to the collision invariants: density, momentum and kinetic energy. However,
one may include higher-order polynomials in this subspace. Their number and
order depend on the physical processes that one wishes to treat
\textquotedblleft exactly\textquotedblright . Thus, one may take into
account non-invariant values like the pressure tensor and heat flux. The
projection operator $\widehat{N}$ maps the state vector onto the remaining
\textquotedblleft non-hydrodynamic subspace\textquotedblright . Since we are
interested in a model operator describing the first 13 moments correctly we
take the operator $\widehat{H}$ in the following form:

{%
\begin{equation}
\widehat{H}=\sum_{i=1}^{13}\left\vert \Psi _{i}\right\rangle \left\langle
\Psi _{i}\right\vert ,  \label{b.1}
\end{equation}%
} where the first 13 Hermite polynomials in a Cartesian coordinate system
are \cite{KliStat}:

{$\left\vert \Psi _{1}\right\rangle =1;$ $\left\vert \Psi
_{r+1}\right\rangle =u_{r},$ $r=1,2,3;$ $\left\vert \Psi _{5}\right\rangle =%
\sqrt{1/6}(u^{2}-3);$ }$\left\vert \Psi _{6}\right\rangle =\sqrt{3/2}%
(u_{x}^{2}-\frac{1}{3}u^{2});$ $\left\vert \Psi _{7}\right\rangle
=1/2(u_{y}^{2}-u_{z}^{2});$ $\left\vert \Psi _{8}\right\rangle =u_{x}u_{z};$ 
$\left\vert \Psi _{9}\right\rangle =u_{x}u_{y};$ $\left\vert \Psi
_{10}\right\rangle =u_{y}u_{z};$ $\left\vert \Psi _{r+10}\right\rangle ={%
\sqrt{1/10}u_{r}(u^{2}-5),}$ ${r=1,2,3}$,

$\mathbf{u}=(\mathbf{p}-m\mathbf{V})/(mT)^{1/2}$ is the dimensionless
velocity.

The linearized collision operator is:{%
\begin{equation}
\delta \widehat{I}=\widehat{H}\delta \widehat{I}\widehat{H}+\widehat{H}%
\delta \widehat{I}\widehat{N}+\widehat{N}\delta \widehat{I}\widehat{H}+%
\widehat{N}\delta \widehat{I}\widehat{N}.  \label{b.2}
\end{equation}%
}\ Since the first five Hermite polynomials are the eigenfunctions of the
collision operator for identical particles corresponding to the zero
eigenvalue, it follows that for $1\leq i\leq 5${\ one has $\widehat{H}\delta 
\widehat{I}\widehat{H}=\widehat{H}\delta \widehat{I}\widehat{N}=\widehat{N}%
\delta \widehat{I}\widehat{H}=0.$ }The higher Hermite polynomials are
eigenfunctions of the collision operator only for a{\ Maxwell's molecule}
with a model repulsive potential proportional to $r^{-4}$. In this case,
non-diagonal matrix elements $\widehat{H}\delta \widehat{I}\widehat{N}$\ and 
$\widehat{N}\delta \widehat{I}\widehat{H}$\ vanish:{\ 
\begin{equation}
\widehat{H}\delta \widehat{I}\widehat{N}=\widehat{N}\delta \widehat{I}%
\widehat{H}=0.  \label{d.4}
\end{equation}%
} For any other interaction potentials the Hermite polynomials are not the
eigenfunctions of the collision operator, the equality (\ref{d.4}) does not
hold and the collision operator matrix elements contain non-diagonal
elements. Our first approximation is that we accept (\ref{d.4}) as a valid
formula for the Boltzmann gas of hard spheres and Coulomb plasma. However,
the first approximation is not sufficient for describing real gas and
plasma. In the second approximation we take into account only the
non-diagonal terms closest to the diagonal. As we will show below, the
corrections for a Boltzmann gas of hard spheres turns out to be small, but
for Coulomb systems they are not small and comparable with the Spitzer
corrections to transport coefficients. We can continue this process and take
into account in the third approximation the next, non-diagonal, terms more
distant from the diagonal elements. We performed these calculations and
found that the third approximation yields very small corrections (compared
to the second approximation), that can be neglected.

Since for one-component systems the operator is Hermitian and isotropic,
Wigner-Ekkart theorem \cite{W-E} implies{\ and }the selection rule follows:
the contribution to the non-diagonal matrix elements $\widehat{H}\delta 
\widehat{I}\widehat{N}$\ and $\widehat{N}\delta \widehat{I}\widehat{H}$ is
given only by polynomials with identical pairs of orbital numbers. For
example, for the polynomial $\left\vert \Psi _{6}\right\rangle =\frac{\sqrt{3%
}}{2}(u_{x}u_{x}-\frac{1}{3}u^{2})$\ defining the $xx$\ component of the
pressure tensor, the non-zero contribution to the non-diagonal matrix
elements is given by non-hydrodynamic polynomials of higher order in $u^{2}$
but with the same values of $l$\ and $\ m$\ ($l=2;m=2$). $\Psi _{6}^{(2)}=%
\sqrt{\frac{3}{14}}\frac{1}{2}(u^{2}-7)(u_{x}u_{x}-\frac{1}{3}u^{2})$\.{ }

The main modeling procedure consists of approximating the non-hydrodinamic
contribution. If the operator $\widehat{H}$\ involves the first 13 Hermite
polynomials, then the neglect of the term $\widehat{N}\delta \widehat{I}%
\widehat{N}$ does not affect calculations for such transport coefficients as
viscosity and heat conductivity. Nevertheless the approximation{\ 
\begin{equation}
\widehat{N}\delta \widehat{I}\widehat{N}=-\nu \widehat{N}  \label{b.6}
\end{equation}%
which is reduced to the partial breach of the fine structure of its spectrum
by the contraction of all the eigenvalues of the }$\widehat{N}$ to the
minimum value, {allows one to describe at least qualitatively the 'tails' of
neglected 'non-hydrodynamic' terms ($\nu $\ corresponds to the longest
non-hydrodynamic relaxation time). An account of these 'tails' may be
important at the kinetic level of fluctuation description. Using this
approximation one may rewrite, in a first approximation, that corresponds to
the Maxwell's molecule, the model operator as follows:{\ 
\begin{equation}
\delta \widehat{I}=-\nu \widehat{Id}+\widehat{H}(\delta \widehat{I}+\nu )%
\widehat{H}.  \label{b.8}
\end{equation}%
}}

{{For the 13 moment basis for $\widehat{H}$\ in the first approximation one
has\ {\ 
\begin{equation}
\delta \widehat{I}\left\vert h\right\rangle =-\nu \left\vert h\right\rangle
+\nu \sum_{i=1}^{5}\left\vert \Psi _{i}\right\rangle \left\langle \Psi
_{i}|h\right\rangle +\sum_{i=6}^{13}\left\vert \Psi _{i}\right\rangle
(\left\langle \Psi _{i}\right\vert \delta \widehat{I}\left\vert \Psi
_{i}\right\rangle +\nu )\left\langle \Psi _{i}|h\right\rangle .  \label{b.9}
\end{equation}%
In the same approximation, the expression for the intensity of the Langevin
source has the form:}}} 
\begin{equation}
(yy)_{\omega ,\mathbf{k,p}_{1},\mathbf{p}_{2}}=2f^{0}(\mathbf{p}_{1})[\nu {%
\delta (}\mathbf{p}_{1}-\mathbf{p}_{2})-\nu f^{0}(\mathbf{p}%
_{2})\sum_{i=1}^{5}\Psi _{i}(\mathbf{u}_{1})\Psi _{i}(\mathbf{u}_{2})+f^{0}(%
\mathbf{p}_{2})\sum_{i=6}^{13}\Psi _{i}(\mathbf{u}_{1})\Psi _{i}(\mathbf{u}%
_{2})(\left\langle \Psi _{i}\right\vert \delta \widehat{I}\left\vert \Psi
_{i}\right\rangle +\nu )]  \label{b.7}
\end{equation}%
The first two terms in (\ref{b.7}) correspond to the BGK model (\ref{a.14}).

{{{In the second approximation, {the nearest non-diagonal entries appear }in
(\ref{b.9}): 
\begin{equation}
\sum_{i=6}^{13}\left\vert \Psi _{i}\right\rangle \left\langle \Psi
_{i}\right\vert \delta \widehat{I}\left\vert \Psi _{i}^{(2)}\right\rangle
\left\langle \Psi _{i}^{(2)}|h\right\rangle ,  \label{b.9b}
\end{equation}%
where the non-hydrodynamic polynomials, which we take into account, are 
\begin{equation}
\Psi _{i}^{(2)}(u)=\frac{1}{\sqrt{14}}(u^{2}-7)\Psi _{i}(u),\ 6\leq i\leq 10
\label{b.9c}
\end{equation}%
\begin{equation}
\Psi _{r+10}^{(2)}(u)=\frac{1}{\sqrt{280}}(u^{4}-14u^{2}+35)\Psi _{r}(u),\
1\leq r\leq 3.  \label{b.9a}
\end{equation}%
The non-hydrodynamical moments of polynomials ({{\ref{b.9c}}, }\ref{b.9a})
are determined in the Fourier presentation and }}}${k=0}$ by the following
equations{{{%
\begin{equation}
(-i\omega -\left\langle \Psi _{i}^{(2)}\right\vert \delta \widehat{I}%
\left\vert \Psi _{i}^{(2)}\right\rangle )\left\langle \Psi
_{i}^{(2)}|h\right\rangle _{\omega }=\left\langle \Psi _{i}^{(2)}\right\vert
\delta \widehat{I}\left\vert \Psi _{i}\right\rangle \left\langle \Psi
_{i}|h\right\rangle _{\omega }.  \label{b.10}
\end{equation}%
}Thus in the second approximation, the linearized model collision operator
has the form 
\begin{equation}
\delta \widehat{I}\left\vert h\right\rangle _{\omega }=-\nu \left\vert
h\right\rangle _{\omega }+\nu \sum_{i=1}^{5}\left\vert \Psi
_{i}\right\rangle \left\langle \Psi _{i}|h\right\rangle _{\omega
}-\sum_{i=6}^{13}\left\vert \Psi _{i}\right\rangle (_{i}\Lambda
_{i}^{(2)}(\omega )-\nu )\left\langle \Psi _{i}|h\right\rangle _{\omega },
\label{b.11}
\end{equation}%
where{\ 
\begin{equation}
_{i}\Lambda _{i}^{(2)}(\omega )=-\left\langle \Psi _{i}\right\vert \delta 
\widehat{I}\left\vert \Psi _{i}\right\rangle -\frac{\left\langle \Psi
_{i}\right\vert \delta \widehat{I}\left\vert \Psi _{i}^{(2)}\right\rangle
^{2}}{-i\omega -\left\langle \Psi _{i}^{(2)}\right\vert \delta \widehat{I}%
\left\vert \Psi _{i}^{(2)}\right\rangle }  \label{b.13}
\end{equation}%
contains the square of the non-diagonal entries and the projection of the {%
resolvent of the }kinetic equation onto the non-hydrodynamical subspace.
Here we take into account non-stationarity of non-hydrodynamical moments.
Thus, although the original collision integral is Markov, the part projected
onto the subspace of the 13 moments becomes, in the second approximation, a
frequency dependent operator. A similar situation occurs in
quantum-mechanical perturbation theory. Note that in the Markov
approximation, the second order corrections in {{(\ref{b.13}) }}are negative
for any interaction potentials. }}}Calculate now matrix elements of the
operator for a special interaction potential, namely for the Coulomb plasma
and the Boltzmann hard sphere gas.

\subsubsection{Coulomb One-Component Plasma}

For the Coulomb one-component plasma, the time evolution of the fluctuations
of the particle distribution function is determined by the equation

{%
\begin{equation}
\frac{\partial }{\partial t}\left\vert h\right\rangle +\widehat{\Phi }%
\left\vert h\right\rangle +\widehat{V}\left\vert h\right\rangle -\delta 
\widehat{I}\left\vert h\right\rangle =\left\vert y\right\rangle ,
\label{c.1}
\end{equation}%
}

where $\widehat{\Phi }$ is the operator that describes the free motion, and $%
\widehat{V}$\ is the linearized Vlasov operator. The intensity of the
Langevin source $\left\vert y\right\rangle $ in {{{(\ref{c.1}) differs from
the expressions {{(\ref{a.14})}} and {{(\ref{b.7})}} by the quantity }}}$%
[1/f^{0}(\mathbf{p}_{1})f^{0}(\mathbf{p}_{2})]$. $\delta \widehat{I}$ is the
linearized collision operator, its action being determined by \cite{Balu}

{%
\begin{equation}
\delta \widehat{I}\left\vert h\right\rangle =\frac{1}{f^{0}}\int (\frac{%
\partial }{\partial p_{r}}-\frac{\partial }{\partial p_{r}^{\prime }})Q^{rs}(%
\frac{\partial }{\partial p_{s}}-\frac{\partial }{\partial p_{s}^{\prime }}%
)f^{0}(\mathbf{p})f^{0}(\mathbf{p}^{\prime })\left( h(\mathbf{p})+h(\mathbf{p%
}^{\prime })\right) d\mathbf{p}^{\prime },  \label{c.2}
\end{equation}%
}

where the kernel $Q^{rs}$ in {the Balescu-Lenard form} is{%
\begin{equation*}
Q^{rs}=2e^{4}\int d\mathbf{k}\frac{k_{r}k_{s}}{k^{4}\left\vert \varepsilon (%
\mathbf{kv,k})\right\vert ^{2}}\delta (\mathbf{kv-kv}^{\prime }).
\end{equation*}%
}

The matrix element of the operator {{{(\ref{c.2}) can be represented in the
form}}}

{%
\begin{equation}
\left\langle \Psi _{i}\right\vert \delta \widehat{I}\left\vert \Psi
_{j}\right\rangle =-\frac{1}{2}\int Q^{rs}f^{0}(\mathbf{p})f^{0}(\mathbf{p}%
^{\prime })(\frac{\partial \Psi _{i}}{\partial \mathbf{p}_{r}}-\frac{%
\partial \Psi _{i}}{\partial \mathbf{p}_{r}^{\prime }})(\frac{\partial \Psi
_{j}}{\partial \mathbf{p}_{s}}-\frac{\partial \Psi _{j}}{\partial \mathbf{p}%
_{s}^{\prime }})d\mathbf{p}d\mathbf{p}^{\prime }.  \label{c.5}
\end{equation}%
}For $i=j$, $\left\langle \Psi _{i}\right\vert \delta \widehat{I}\left\vert
\Psi _{j}\right\rangle \leq 0$. Vanishing the matrix elements $\left\langle
\Psi _{i}\right\vert \delta \widehat{I}\left\vert \Psi _{j}\right\rangle $
(for $i=1-5$) corresponds to the fivefold degenerate zero eigenvalue of the
operator $\delta \widehat{I}$. It is easy to show that the matrix elements
of $\widehat{H}\delta \widehat{I}\widehat{H}$ have the values

\begin{equation}
\left\langle \Psi _{i}\right\vert {\ \delta }\widehat{I}\left\vert \Psi
_{j}\right\rangle =-{\ \delta }_{ij}[{\ \Lambda }_{1}\sum_{k=6}^{10}{\
\delta }_{ik}+{\ \Lambda }_{2}\sum_{k=11}^{13}{\ \delta }_{ik}],  \label{c.6}
\end{equation}%
where $\Lambda _{1},$ $\Lambda _{2}$ are the reciprocal relaxation times of
the pressure tensor and the heat flux vector which in terms of the plasma
parameters are as follows: 
\begin{equation}
{\ \Lambda }_{1}=\frac{1}{10}\frac{\lambda }{\pi ^{3/2}}\omega _{L}\ln \frac{%
1}{\lambda };\ {\ \Lambda }_{2}=\frac{2}{3}{\ \Lambda }_{1},  \label{c.7}
\end{equation}%
where $\omega _{L}=\left( \frac{4\pi ne^{2}}{m}\right) ^{1/2}$ is the
electron plasma frequency, and $\lambda =n^{-1}k_{D}^{3}$ is the plasma
parameter.

The matrix elements of the operators $\widehat{N}\delta \widehat{I}\widehat{N%
}$\ and $\widehat{H}\delta \widehat{I}\widehat{N}$ defining 'tails' and the
corrections of the second approximation, have for a Coulomb plasma the
values \cite{Bel-PV}

$\left\langle \Psi _{14}\right\vert {\ \delta }\widehat{I}\left\vert \Psi
_{14}\right\rangle =-\frac{2}{3}{\ \Lambda }_{1};\ \Psi _{14}=\frac{1}{\sqrt{%
120}}(u^{4}-10u^{2}+15),$

$\left\langle \Psi _{15}\right\vert {\ \delta }\widehat{I}\left\vert \Psi
_{15}\right\rangle =...=\left\langle \Psi _{21}\right\vert {\ \delta }%
\widehat{I}\left\vert \Psi _{21}\right\rangle =-\frac{3}{2}{\Lambda }_{1};\
\Psi _{15}=u_{x}u_{y}u_{z},$

$\left\langle \Psi _{22}\right\vert {\ \delta }\widehat{I}\left\vert \Psi
_{22}\right\rangle =...=\left\langle \Psi _{35}\right\vert {\ \delta }%
\widehat{I}\left\vert \Psi _{35}\right\rangle =-\frac{191}{16}{\Lambda }%
_{1}; $ $\Psi _{22}=\frac{1}{\sqrt{105}}\frac{1}{8}%
(35u_{x}^{4}-30u^{2}u_{x}^{2}+3u^{4}),$

$\left\langle \Psi _{i}^{(2)}\right\vert {\delta }\widehat{I}\left\vert \Psi
_{i}^{(2)}\right\rangle =-\frac{201}{168}{\Lambda }_{1};\ \left\langle \Psi
_{i}\right\vert {\delta }\widehat{I}\left\vert \Psi _{i}^{(2)}\right\rangle =%
\frac{3}{2\sqrt{14}}{\Lambda }_{1};\ 6\leq i\leq 10,$

$\left\langle \Psi _{10+r}^{(2)}\right\vert {\delta }\widehat{I}\left\vert
\Psi _{10+r^{\prime }}^{(2)}\right\rangle =-\frac{15}{14}{\delta }%
_{rr^{\prime }}{\Lambda }_{1},$

$\left\langle \Psi _{10+r}\right\vert {\delta }\widehat{I}\left\vert \Psi
_{10+r^{\prime }}^{(2)}\right\rangle =\frac{1}{7}{\delta }_{rr^{\prime }}{%
\Lambda }_{1};\ {1}\leq r\leq 3.$

This estimate implies that the higher tensor character of the polynomial
leads to higher values of the matrix elements and for polynomials with the
same tensor character the matrix elements are greater for polynomials of
higher order in $u^{2}$. The smallest value of the diagonal matrix elements
for the non-hydrodynamical polynomial is achieved for the $\Psi _{14}$\
polynomial and equals \ $\Lambda _{2}$\ (the heat flux relaxation
frequency). The same holds for Maxwell's molecule and the Boltzmann gas hard
sphere at least. Thus in the first approximation the linearized collision
operator\ and the intensity of the Langevin source have the form:{\ 
\begin{equation}
\delta \widehat{I}\left\vert h\right\rangle =-\frac{2}{3}\Lambda
_{1}\left\vert h\right\rangle +\frac{2}{3}\Lambda
_{1}\sum_{i=1}^{5}\left\vert \Psi _{i}\right\rangle \left\langle \Psi
_{i}|h\right\rangle -\frac{1}{3}\Lambda _{1}\sum_{i=6}^{10}\left\vert \Psi
_{i}\right\rangle \left\langle \Psi _{i}|h\right\rangle ,  \label{b.25}
\end{equation}%
}%
\begin{equation}
(yy)_{\omega ,\mathbf{k,p}_{1},\mathbf{p}_{2}}=\frac{4}{3}\Lambda _{1}f^{0}(%
\mathbf{p}_{1})[{\delta (}\mathbf{p}_{1}-\mathbf{p}_{2})-f^{0}(\mathbf{p}%
_{2})\sum_{i=1}^{5}\Psi _{i}(\mathbf{u}_{1})\Psi _{i}(\mathbf{u}_{2})+\frac{1%
}{2}f^{0}(\mathbf{p}_{2})\sum_{i=6}^{13}\Psi _{i}(\mathbf{u}_{1})\Psi _{i}(%
\mathbf{u}_{2})].  \label{b.26}
\end{equation}%
{We see that the polynomials corresponding to the heat flux disappear. It is
easy to show that {{(\ref{b.25}) and (\ref{b.26})}} gives the well-known
values of the transport coefficients in the hydrodynamic equations and the
Landau-Lifshitz formulas \cite{Landau}.}

{In the second approximation it holds{\ 
\begin{equation}
\delta \widehat{I}\left\vert h\right\rangle _{\omega }=-\nu (\left\vert
h\right\rangle _{\omega }-\sum_{i=1}^{13}\left\vert \Psi _{i}\right\rangle
\left\langle \Psi _{i}|h\right\rangle _{\omega })-\sum_{i=6}^{10}\Lambda
_{1}^{(2)}(\omega )\left\vert \Psi _{i}\right\rangle \left\langle \Psi
_{i}|h\right\rangle _{\omega }-\sum_{i=11}^{13}\Lambda _{2}^{(2)}(\omega
)\left\vert \Psi _{i}\right\rangle \left\langle \Psi _{i}|h\right\rangle
_{\omega },  \label{b.25a}
\end{equation}%
and}}%
\begin{equation*}
(yy)_{\omega ,\mathbf{k,p}_{1},\mathbf{p}_{2}}=2f^{0}(\mathbf{p}_{1})[\nu {%
\delta (}\mathbf{p}_{1}-\mathbf{p}_{2})-\nu f^{0}(\mathbf{p}%
_{2})\sum_{i=1}^{5}\Psi _{i}(\mathbf{u}_{1})\Psi _{i}(\mathbf{u}_{2})
\end{equation*}%
\begin{equation}
+\mbox{Re}\Lambda _{1}^{(2)}(\omega )f^{0}(\mathbf{p}_{2})\sum_{i=6}^{10}%
\Psi _{i}(\mathbf{u}_{1})\Psi _{i}(\mathbf{u}_{2})+\mbox{Re}\Lambda
_{2}^{(2)}(\omega )f^{0}(\mathbf{p}_{2})\sum_{i=11}^{13}\Psi _{i}(\mathbf{u}%
_{1})\Psi _{i}(\mathbf{u}_{2})]\text{,}  \label{b.27}
\end{equation}

{where}%
\begin{equation*}
\text{ }{\Lambda }_{1}^{(2)}{\ (\omega )=\Lambda }_{1}(1-\frac{{\ \Lambda }%
_{1}9/56}{-i{\ \omega +\Lambda }_{1}205/168});{\ }
\end{equation*}%
{{\ 
\begin{equation}
{\ \Lambda }_{2}^{(2)}{\ (\omega )=\Lambda }_{2}(1-\frac{{\ \Lambda }_{1}3/14%
}{-i{\ \omega +\Lambda }_{1}15/14}).  \label{b.28}
\end{equation}%
In the Markov approximation ($\omega =0$) 
\begin{equation}
{\ \Lambda }_{1}^{(2)}{\ =\Lambda }_{1}(1-\frac{27}{205});\ {\ \Lambda }%
_{2}^{(2)}{\ =\Lambda }_{2}(1-\frac{1}{5}).  \label{b.29}
\end{equation}%
{\ the corrections are rather significant and comparable with the Spitzer
corrections {\cite{Spitzer}}}. In the third approximation the relaxation
frequencies vary by no more than one per cent.}}

From the scalar product of {{{\ (\ref{c.1})}}} with $\left\vert \Psi
_{k}\right\rangle $ $(1\leq k\leq 13)$ there follows a system of equations
for the hydrodynamic moments

\begin{equation*}
\frac{\partial }{\partial t}{\delta n+n}\frac{\partial }{\partial \mathbf{q}}%
{\delta }\mathbf{V=}0\mathbf{;}\text{ }
\end{equation*}%
\begin{equation*}
\text{ }\frac{\partial }{\partial t}n{\delta }V_{i}\mathbf{=-}\frac{1}{m}%
\frac{\partial {\delta (nT)}}{\partial q_{i}}\mathbf{-}\frac{1}{m}\frac{%
\partial {\delta P}_{ij}}{\partial q_{j}}\mathbf{-}\frac{1}{m}{\delta E}_{i};%
\text{ }
\end{equation*}%
\begin{equation}
\text{ }\frac{\partial }{\partial t}{\delta }T+\frac{2}{3}\frac{T}{n}\frac{%
\partial }{\partial \mathbf{q}}{\delta }\mathbf{V+}\frac{2}{3}\frac{1}{n}%
\frac{\partial }{\partial \mathbf{q}}{\delta }\mathbf{S=}0;  \label{b.30}
\end{equation}

\begin{equation*}
\frac{\partial }{\partial t}{\delta P}_{ij}-nT\left( \frac{\partial {\delta V%
}_{i}}{\partial q_{j}}+\frac{\partial {\delta V}_{j}}{\partial q_{i}}-\frac{2%
}{3}{\delta }_{ij}\frac{\partial {\delta V}_{k}}{\partial q_{k}}\right) -%
\frac{2}{5}\left( \frac{\partial {\delta S}_{i}}{\partial q_{j}}+\frac{%
\partial {\delta S}_{j}}{\partial q_{i}}-\frac{2}{3}{\delta }_{ij}\frac{%
\partial {\delta S}_{k}}{\partial q_{k}}\right) \text{ }
\end{equation*}%
\begin{equation}
\mathbf{=-}\Lambda _{1}^{(2)}(\omega ){\delta P}_{ij}+\frac{1}{m}\langle
(p_{i}p_{j}-\frac{1}{3}{\delta }_{ij}p^{2})\left\vert y\right\rangle \mathbf{%
;}\text{ }  \label{b.30b}
\end{equation}%
\bigskip

\begin{equation}
\frac{\partial }{\partial t}{\delta S}_{i}+\frac{5}{2}\frac{nT}{m}\frac{%
\partial {\delta T}}{\partial q_{i}}-\frac{T}{m}\frac{\partial {\delta P}%
_{ij}}{\partial q_{j}}=-\Lambda _{2}^{(2)}(\omega ){\delta S}_{i}+\frac{1}{%
2m^{2}}\langle p_{i}(p^{2}-5Tm)\left\vert y\right\rangle ,  \label{b.30c}
\end{equation}%
where%
\begin{equation*}
{\delta P}_{ij}=nT\{{\delta }_{ix}{\delta }_{jx}\sqrt{\frac{4}{3}}\langle
\Psi _{6}\left\vert h\right\rangle +{\delta }_{iy}{\delta }_{jy}(\langle
\Psi _{7}\left\vert h\right\rangle )-\frac{1}{\sqrt{3}}\langle \Psi
_{6}\left\vert h\right\rangle )-{\delta }_{iz}{\delta }_{jz}(\langle \Psi
_{7}\left\vert h\right\rangle )+\frac{1}{\sqrt{3}}\langle \Psi
_{6}\left\vert h\right\rangle )+({\delta }_{ix}{\delta }_{jz}+{\delta }_{iz}{%
\delta }_{jx})\langle \Psi _{8}\left\vert h\right\rangle
\end{equation*}%
\begin{equation}
+({\delta }_{ix}{\delta }_{jy}+{\delta }_{iy}{\delta }_{jx})\langle \Psi
_{9}\left\vert h\right\rangle +({\delta }_{iy}{\delta }_{jz}+{\delta }_{iz}{%
\delta }_{jy})\langle \Psi _{10}\left\vert h\right\rangle \}  \label{b.30m}
\end{equation}%
is the fluctuation of the{\ pressure tensor and}%
\begin{equation}
{\delta S}_{r}=\frac{mn}{2}(\frac{T}{2})^{3/2}\sqrt{10}\langle \Psi
_{10+r}\left\vert h\right\rangle  \label{b.30n}
\end{equation}%
is the fluctuation of the heat flux.

In deriving {{{(\ref{b.30b}) and (\ref{b.30c}), we have retained the terms
with time derivatives, which for a plasma do not vanish in the
long-wavelength approximation }}}$k\rightarrow 0$. Making a Fourier
transformation of {{{(\ref{b.30b}), we obtain}}}

\begin{equation}
{\delta P}_{ij}(\omega )={\delta \pi }_{ij}(\omega )+\frac{1}{-i\omega
+\Lambda _{1}^{(2)}(\omega )}\{nT\left( \frac{\partial {\delta V}_{i}}{%
\partial q_{j}}+\frac{\partial {\delta V}_{j}}{\partial q_{i}}-\frac{2}{3}{%
\delta }_{ij}\frac{\partial {\delta V}_{k}}{\partial q_{k}}\right) +\frac{2}{%
5}\left( \frac{\partial {\delta S}_{i}}{\partial q_{j}}+\frac{\partial {%
\delta S}_{j}}{\partial q_{i}}-\frac{2}{3}{\delta }_{ij}\frac{\partial {%
\delta S}_{k}}{\partial q_{k}}\right) \},\text{ }  \label{b.30d}
\end{equation}%
where%
\begin{equation}
{\delta \pi }_{ij}(\omega )=\frac{1}{-i\omega +\Lambda _{1}^{(2)}(\omega )}%
\frac{1}{m}\langle (p_{i}p_{j}-\frac{1}{3}{\delta }_{ij}p^{2})\left\vert
y\right\rangle .  \label{b.30e}
\end{equation}

Similarly, it follows from {{{(\ref{b.30c}) that}}}%
\begin{equation}
{\delta S}_{i}(\omega )=\sigma _{i}(\omega )-\frac{5}{2m}\frac{nT}{-i\omega
+\Lambda _{2}^{(2)}(\omega )}\frac{\partial }{\partial q_{i}}{\delta T}%
(\omega ),  \label{b.30f}
\end{equation}%
where%
\begin{equation}
\sigma _{i}(\omega )=-\frac{1}{-i\omega +\Lambda _{2}^{(2)}(\omega )}\frac{1%
}{2m^{2}}\langle p_{i}(p^{2}-5Tm)\left\vert y\right\rangle .  \label{b.30g}
\end{equation}

Using the expression for the spectral function of the Langevin source in the
kinetic equation {{{(\ref{b.27}), we obtain}}}

\begin{equation}
({\delta \pi }_{ij}{\delta \pi }_{rl})_{\omega }=({\delta }_{ir}{\delta }%
_{jl}+{\delta }_{il}{\delta }_{jr}-\frac{2}{3}{\delta }_{ij}{\delta }_{rl})%
\frac{2nT^{2}\mbox{Re}\Lambda _{1}^{(2)}(\omega )}{\left\vert -i\omega
+\Lambda _{1}^{(2)}(\omega )\right\vert ^{2}}.\text{ }  \label{b.30h}
\end{equation}%
Introducing the frequency-dependent viscosity%
\begin{equation}
\eta (\omega )=\frac{nT}{-i\omega +\Lambda _{1}^{(2)}(\omega )}
\label{b.30j}
\end{equation}%
we can rewrite the expression {{{(\ref{b.30h}) as}}}%
\begin{equation}
({\delta \pi }_{ij}{\delta \pi }_{rl})_{\omega }=2T\mbox{Re}\eta (\omega )({%
\delta }_{ir}{\delta }_{jl}+{\delta }_{il}{\delta }_{jr}-\frac{2}{3}{\delta }%
_{ij}{\delta }_{rl}).  \label{b.30k}
\end{equation}%
Similarly,%
\begin{equation}
(\sigma _{r}\sigma _{l})_{\omega }=2\frac{T^{2}}{\kappa }{\delta }_{rl}%
\mbox{Re}\lambda (\omega ),  \label{b.30l}
\end{equation}%
with frequency-dependent thermal conductivity%
\begin{equation}
\lambda (\omega )=\frac{5}{2}\frac{nT\kappa }{m}\frac{1}{-i\omega +\Lambda
_{2}^{(2)}(\omega )}.  \label{b.30z}
\end{equation}%
In the low-frequency limit $\omega \ll \Lambda _{1},\Lambda _{2}$ Eqs. {{{(%
\ref{b.30k}, {{\ref{b.30l}}}) yield the classical results }}}\cite{Landau}.
In the first approximation ($\Lambda _{1,2}^{(2)}(\omega )=\Lambda _{1,2}$),
the results {{{(\ref{b.30k}, {{\ref{b.30l}}}) go over into the results
obtained earlier }}}\cite{Rosmus}. In the general case, when the non-Markov
processes are important, the transport coefficients {{{(\ref{b.30j}) and (%
\ref{b.30z}) that occur in the expressions (\ref{b.30k}) and (\ref{b.30l})
contain continued fractions.}}}

\subsubsection{Boltzmann Gas of Hard Spheres}

For a Boltzmann gas of hard spheres, the linearized collision operator has
the form

{%
\begin{equation}
\delta \widehat{I}\left\vert h\right\rangle =\frac{R^{2}}{2f^{0}(\mathbf{p})}%
\int \left\vert \mathbf{(Ve)}\right\vert f^{0}(\mathbf{p})f^{0}(\mathbf{p}%
_{1})\times \lbrack h(\mathbf{p}_{1}^{\prime })+h(\mathbf{p}^{\prime })-h(%
\mathbf{p}_{1})-h(\mathbf{p})]d\mathbf{p}_{1}d\Omega ,  \label{d.31a}
\end{equation}%
}where $R$ is the interaction range, $\mathbf{V=v-v}_{1}$ is the relative
velocity of the particles, $\mathbf{v}^{\prime }=\mathbf{v+e(eV),}$ $\mathbf{%
v}_{1}^{\prime }=\mathbf{v}_{1}\mathbf{-e(eV),}$ $\mathbf{e}$ is an
arbitrary unit vector.

The matrix elements of the operator {{{(\ref{d.31a}) have the form}}\ } 
\begin{equation}
\left\langle \Psi _{i}\right\vert \delta \widehat{I}\left\vert \Psi
_{j}\right\rangle =-\frac{R^{2}}{8n}\int \left\vert \mathbf{Ve}\right\vert
f^{0}(\mathbf{p})f^{0}(\mathbf{p}^{\prime })[\Psi _{i}(\mathbf{p}%
_{1}^{\prime })+\Psi _{i}(\mathbf{p}^{\prime })-\Psi _{i}(\mathbf{p}%
_{1})-\Psi _{i}(\mathbf{p})][\Psi _{j}(\mathbf{p}_{1}^{\prime })+\Psi _{j}(%
\mathbf{p}^{\prime })-\Psi _{j}(\mathbf{p}_{1})-\Psi _{j}(\mathbf{p})]d%
\mathbf{p}d\mathbf{p}_{1}.  \label{b.31b}
\end{equation}%
In the hydrodynamic space

{%
\begin{equation}
\left\langle \Psi _{i}\right\vert \delta \widehat{I}\left\vert \Psi
_{j}\right\rangle =-\delta _{ij}[{\Lambda }_{1}^{G}\sum_{k=6}^{10}\delta
_{ik}+{\Lambda }_{2}^{G}\sum_{k=11}^{13}\delta _{ik}],  \label{b.31c}
\end{equation}%
}where

\begin{equation}
{\Lambda }_{1}^{G}=\frac{16}{5}\sqrt{\pi }nR^{2}\sqrt{\frac{T}{m}},\text{ }{%
\Lambda }_{2}^{G}=\frac{32}{15}\sqrt{\pi }nR^{2}\sqrt{\frac{T}{m}}
\label{b.31d}
\end{equation}%
The matrix elements that determine the correction of the second
approximation for the Boltzmann gas take the values

$\left\langle \Psi _{14}\right\vert {\ \delta }\widehat{I}\left\vert \Psi
_{14}\right\rangle =-\frac{2}{3}{\ \Lambda }_{1}^{G};\ \left\langle \Psi
_{15}\right\vert {\ \delta }\widehat{I}\left\vert \Psi _{15}\right\rangle
=...=\left\langle \Psi _{21}\right\vert {\ \delta }\widehat{I}\left\vert
\Psi _{21}\right\rangle =-\frac{3}{2}{\ \Lambda }_{1}^{G},$

$\left\langle \Psi _{i}^{(2)}\right\vert {\delta }\widehat{I}\left\vert \Psi
_{i}^{(2)}\right\rangle =-\frac{17}{14}{\Lambda }_{1}^{G};\ \left\langle
\Psi _{i}\right\vert {\delta }\widehat{I}\left\vert \Psi
_{i}^{(2)}\right\rangle =-\frac{1}{2\sqrt{14}}{\Lambda }_{1}^{G};\ 6\leq
i\leq 10,$

$\left\langle \Psi _{10+r}^{(2)}\right\vert {\delta }\widehat{I}\left\vert
\Psi _{10+r^{\prime }}^{(2)}\right\rangle =-\frac{15}{14}{\delta }%
_{rr^{\prime }}{\Lambda }_{1}^{G},$

$\left\langle \Psi _{10+r}\right\vert {\delta }\widehat{I}\left\vert \Psi
_{10+r^{\prime }}^{(2)}\right\rangle =\frac{1}{3\sqrt{7}}{\delta }%
_{rr^{\prime }}{\Lambda }_{1}^{G};\ 1\leq r\leq 3.$

The second-order corrections for Boltzmann gas are one order less than those
for the Coulomb plasma and one may stop at the first approximation. Thus in
the case of a Boltzmann gas of hard spheres the collision operator may be
represented in the form {{{\cite{Bel}}}}

{%
\begin{equation}
I\{f\}=-\nu \{f-f^{0}(1-P_{ij}\frac{{\delta v}_{i}{\delta v}_{j}}{4PT}m)\},
\label{eqn37}
\end{equation}%
}Here, we have used the fact that ${\Lambda }_{2}^{G}=2/3{\Lambda }%
_{1}^{G}=\nu .$

For arbitrary Prandtl number $\Pr $ our model takes the form:{%
\begin{equation}
I\{f\}=-\nu \left\{ f-f^{0}\left[ 1-\left( \frac{1-\Pr }{\Pr }\right) P_{ij}%
\frac{{\delta v}_{i}{\delta v}_{j}}{2PT}m\right] \right\} ,  \label{eqn37a}
\end{equation}%
}

{In the equilibrium state, $P_{ij}=0$\ and $I\{f^{0}\}=0$. In the equation
for the heat flux only the first term in {{(\ref{eqn37a})}} contributes. The
relaxation of the pressure tensor is determined by both the first and the
last terms in this equation.}

Now we will prove the $H$ theorem for our model {{{(\ref{eqn37a}):}}}

\begin{equation}
\frac{\partial }{\partial t}H(t)=\nu \int (f-\Phi )\log \frac{f}{\Phi }d%
\mathbf{p+}\nu \int (f-\Phi )\left\{ \log {f}^{0}+\log \left[ 1-\left( \frac{%
1-\Pr }{\Pr }\right) P_{kl}\frac{{\delta v}_{k}{\delta v}_{l}}{2PT}m\right]
\right\} d\mathbf{p},  \label{eqn38a}
\end{equation}%
where $\Phi =f^{0}\left[ 1-\left( \frac{1-\Pr }{\Pr }\right) P_{ij}\frac{{%
\delta v}_{i}{\delta v}_{j}}{2PT}m\right] .$

The first term in the brace in {{{(\ref{eqn38a})}}} vanishes, while the
second term can be expanded as: $\log (1+x)=x-x^{2}/2+x^{3}/3...;$ for \ $%
-1<x\leq 1.$ \ Here only the first term contributes after the integration.
Therefore:%
\begin{equation}
\frac{\partial }{\partial t}H(t)=\nu \int (f-\Phi )\log \frac{f}{\Phi }d%
\mathbf{p-}\nu \int d\mathbf{p}\left\{ f-f^{0}\left[ 1-\left( \frac{1-\Pr }{%
\Pr }\right) P_{ij}\frac{{\delta v}_{i}{\delta v}_{j}}{2PT}m\right] \right\}
\left( \frac{1-\Pr }{\Pr }\right) P_{kl}\frac{{\delta v}_{k}{\delta v}_{l}}{%
2PT}m.  \label{eqn38b}
\end{equation}%
Taking into account that%
\begin{equation*}
\int {f}^{0}({\delta v}_{i}{\delta v}_{j}-{\delta }_{ij}\frac{{\delta v}^{2}%
}{3})({\delta v}_{k}{\delta v}_{l}-{\delta }_{kl}\frac{{\delta v}^{2}}{3})d%
\mathbf{p}=n\frac{T^{2}}{m^{2}}({\delta }_{ik}{\delta }_{jl}+{\delta }_{il}{%
\delta }_{jk}-\frac{2}{3}{\delta }_{ij}{\delta }_{kl}),
\end{equation*}%
we obtain the $H$ theorem:%
\begin{equation}
\frac{\partial }{\partial t}H(t)=\nu \int (f-\Phi )\log \frac{f}{\Phi }d%
\mathbf{p-}\frac{\nu }{2PT}\frac{1-\Pr }{\Pr^{2}}P_{ij}P_{ji}\leq \nu \int
(f-\Phi )\log \frac{f}{\Phi }d\mathbf{p\leq 0}\text{.}  \label{eqn39}
\end{equation}

{Thus our collision integral in the form (\ref{eqn37}) possesses all
necessary properties and is free from the drawbacks of the one-component
model of the BGK model mentioned above. The linearized form of (\ref{eqn37})
is congruent with the linearized ellipsoidal statistical model \cite{Hol,
Cerc}. }

Earlier, another model correctly describing the viscosity and thermal
conductivity relaxation was proposed ad hoc \cite{Shahov}:{\ 
\begin{equation}
I\{f\}=-\nu \{f-f^{0}[1-\Pr m\frac{\mathbf{S}{\delta }\mathbf{v}}{PT}(\frac{m%
{\delta v}^{2}}{5T}-1)]\},  \label{eqn40}
\end{equation}%
}

where{%
\begin{equation*}
\mathbf{S}=m\dint {d\mathbf{p}\delta }\mathbf{v}\frac{{\delta }v^{2}}{2}f%
\text{ }
\end{equation*}%
is }the heat flux. But this model does not give a correct description of
non-hydrodynamic 'tails'{\ }

\subsection{Many-component systems}

When the local equilibrium state is achieved, in the many-component systems,
the stage of relaxation of the mean velocities and temperatures ensues and
we can to restrict the hydrodynamical subspace to the first five
polynomials. Using the technique described above for the one-component
system, one may obtain the following expression for the linearized model of
a many-component system in the five-moment approximation:

{{{%
\begin{equation}
\delta \widehat{I}_{a\mathbf{p}}\delta f_{a}(\mathbf{p})=-\nu _{a}\delta
f_{a}(\mathbf{p})+\sum_{j=1}^{5}\nu _{a}f_{a}^{0}(\mathbf{p})\Psi _{j}^{a}(%
\mathbf{p})\int \Psi _{j}^{a}(\mathbf{p}^{\prime })\delta f_{a}(\mathbf{p}%
^{\prime })d\mathbf{p}^{\prime }+\sum_{b}\sum_{i,j=1}^{5}f_{a}^{0}(\mathbf{p}%
)\Psi _{i}^{a}(\mathbf{p})\left\langle \Psi _{i}^{a}\right\vert \delta 
\widehat{I}\left\vert \Psi _{j}^{b}\right\rangle \int \Psi _{j}^{b}(\mathbf{p%
}^{\prime })\delta f_{b}(\mathbf{p}^{\prime })d\mathbf{p}^{\prime },
\label{f.1}
\end{equation}%
}}}where $f_{a}^{0}(\mathbf{p})$ is the local equilibrium distribution
function (with different temperatures and mean velocities), $\nu _{a}$ is
the inverse time of the heat flux relaxation of component $a$, and $%
\left\langle \Psi _{i}^{a}\right\vert \delta \widehat{I}\left\vert \Psi
_{j}^{b}\right\rangle $ represents the matrix elements of the linearized
collision integral of the Balescu-Lenard integral, for example \cite{Balu}:%
\begin{equation*}
\left\langle \Psi _{i}^{a}\right\vert \delta \widehat{I}\left\vert \Psi
_{j}^{b}\right\rangle =-\frac{2}{n_{a}}\sum_{c}\int d\mathbf{k}d\mathbf{p}%
_{1}d\mathbf{p}_{2}\frac{e_{a}^{2}e_{c}^{2}}{k^{4}\left\vert \varepsilon (%
\mathbf{kv}_{1}\mathbf{,k})\right\vert ^{2}}\delta (\mathbf{kv}_{1}-\mathbf{%
kv}_{2})f_{a}^{0}(\mathbf{p}_{1})f_{b}^{0}(\mathbf{p}_{2})
\end{equation*}%
{{{%
\begin{equation}
\times \mathbf{k}\frac{\partial }{\partial \mathbf{p}_{1}}\Psi
_{i}^{a}[\delta _{ab}\mathbf{k}\frac{\partial }{\partial \mathbf{p}_{1}}\Psi
_{j}^{b}-\delta _{bc}\mathbf{k}\frac{\partial }{\partial \mathbf{p}_{2}}\Psi
_{j}^{c}++\left( \frac{\mathbf{k}\delta \mathbf{v}_{2}}{T_{c}}-\frac{\mathbf{%
k}\delta \mathbf{v}_{1}}{T_{a}}\right) \left( \delta _{ab}\Psi _{j}^{b}(%
\mathbf{p}_{1})+\delta _{bc}\Psi _{j}^{c}(\mathbf{p}_{2})\right) ].
\label{f.1a}
\end{equation}%
}}}

In order to recover the form of the model collision integral from its
linearized form {(\ref{f.1}) }it suffices to use the conservation of the
number of elastically interacting particles. This property, as well as the
total momentum and energy conservation, is valid for both mean and
fluctuating quantities. Consequently, the expression for Langevin's source
intensity in the kinetic equation for the fluctuation of the distribution
function {\cite{Lif} }should satisfy the conditions

{{{%
\begin{equation}
\sum_{b}\int \Psi ^{b}(\mathbf{p}_{2})(y_{a}y_{b})_{\omega ,\mathbf{k,p}_{1}%
\mathbf{,p}_{2}}d\mathbf{p}_{2}=0\text{ }  \label{f.3}
\end{equation}%
}}}for $\Psi ^{b}(\mathbf{p}_{2})=1,$ $\mathbf{p}_{2}\mathbf{,}$ $%
p_{2}^{2}/2m_{b}$.

The spectral function of Langevin's source in a non-equilibrium state is
given {\cite{Kogan, GGK, Kli}} by following form: {{{%
\begin{equation}
(y_{a}y_{b})_{\omega ,\mathbf{k,p}_{1}\mathbf{,p}_{2}}=-(\delta \widehat{I}%
_{a\mathbf{p}_{1}}+\delta \widehat{I}_{b\mathbf{p}_{2}})\delta _{ab}\delta (%
\mathbf{p}_{1}-\mathbf{p}_{2})f_{a}(\mathbf{p}_{1})+\delta _{ab}\delta (%
\mathbf{p}_{1}-\mathbf{p}_{2})I_{a}(\mathbf{p}_{1})+I_{ab}(\mathbf{p}_{1},%
\mathbf{p}_{2})\text{ ,}  \label{f.4}
\end{equation}%
}}}where $I_{ab}(\mathbf{p}_{1},\mathbf{p}_{2})$ is the so-called
\textquotedblleft not integrated\textquotedblright\ collision operator {\cite%
{GGK}}:

{{{%
\begin{equation}
\sum_{b}\int I_{ab}(\mathbf{p}_{1},\mathbf{p}_{2})d\mathbf{p}_{2}=I_{a}(%
\mathbf{p}_{1}).  \label{f.4a}
\end{equation}%
}}}

In the case of many-component plasma this \textquotedblleft not
integrated\textquotedblright\ collision operator has the form:

{{{%
\begin{equation}
I_{ab}(\mathbf{p}_{1},\mathbf{p}_{2})=2e_{a}^{2}e_{b}^{2}n_{b}(\frac{%
\partial }{\partial p_{1i}}-\frac{\partial }{\partial p_{2i}})\int \frac{%
k_{i}k_{j}\delta (\mathbf{kv}_{1}\mathbf{-kv}_{2})}{k^{4}\left\vert
\varepsilon (\mathbf{kv}_{1}\mathbf{,k})\right\vert ^{2}}(\frac{\partial }{%
\partial p_{1j}}-\frac{\partial }{\partial p_{2j}})f_{a}f_{b}d\mathbf{k}%
\text{.}  \label{f.5}
\end{equation}%
}}}

Summing {(\ref{f.4})} and {(\ref{f.4a})} over $b$ and integrating over $%
\mathbf{p}_{2}$, and taking into account {(\ref{f.3})}, we get{{{%
\begin{equation}
I_{a}(\mathbf{p}_{1})=\frac{1}{2}\sum_{b}\int (\delta \widehat{I}_{a\mathbf{p%
}_{1}}+\delta \widehat{I}_{b\mathbf{p}_{2}})\delta _{ab}\delta (\mathbf{p}%
_{1}-\mathbf{p}_{2})f_{a}(\mathbf{p}_{1})d\mathbf{p}_{2}.  \label{f.6}
\end{equation}%
}}}Since {(\ref{f.3})} and {(\ref{f.4})} are of a general character, the
relation {(\ref{f.6})} is valid both for \textquotedblleft
exact\textquotedblright\ and model collision integrals. From {(\ref{f.1})}
and {(\ref{f.6})}, we have:

{{{%
\begin{equation}
I_{a}(\mathbf{p})=\frac{1}{2}\sum_{b}\sum_{i=1}^{5}f_{a}^{0}(\mathbf{p})\Psi
_{i}^{a}(\mathbf{p})\left\langle \Psi _{i}^{a}\right\vert \delta \widehat{I}%
\left\vert \Psi _{1}^{b}\right\rangle .  \label{f.6a}
\end{equation}%
The matrix elements }}}$\left\langle \Psi _{i}^{a}\right\vert \delta 
\widehat{I}\left\vert \Psi _{1}^{b}\right\rangle ${{{\ can be calculated for
example for Coulomb plasma%
\begin{equation}
\left\langle \Psi _{i}^{a}\right\vert \delta \widehat{I}\left\vert \Psi
_{1}^{b}\right\rangle =-\sum_{r=1}^{3}{\ \delta }_{ir+1}\sum_{c}\nu _{ac}%
\frac{T_{c}V_{ra}-T_{a}V_{rc}}{T_{a}T_{c}}\sqrt{\frac{m_{a}}{T_{a}}}%
T_{ac}(\delta _{ab}+\delta _{bc})-{\ \delta }_{i5}\sum_{c}\nu _{ac}\sqrt{6}%
\frac{T_{a}-T_{b}}{T_{a}}\frac{\mu _{ac}}{m_{c}}(\delta _{ab}+\delta _{bc}),
\label{f.6b}
\end{equation}%
}}}

where $\nu _{ab}$ is the momentum relaxation frequency for plasma:

{{{%
\begin{equation}
\nu _{ab}=\frac{4}{3}\sqrt{2\pi }\frac{e_{a}^{2}e_{b}^{2}n_{b}\mu _{ab}}{%
m_{a}^{1/2}m_{b}^{3/2}T_{ab}^{3/2}}\ln 1/\lambda  \label{f.8}
\end{equation}%
and 
\begin{equation*}
\mu _{ac}=\frac{m_{a}m_{c}}{m_{a}+m_{c}};T_{ac}=\frac{m_{a}T_{c}+m_{c}T_{a}}{%
m_{a}+m_{c}}
\end{equation*}%
}}}

Substituting (\ref{f.6b}) into (\ref{f.6a}), we obtain a quite simple and,
at the same time, sufficiently rigorous form of the model collision integral
for many-component plasma in the local equilibrium state, which describes in
the customary form the mean velocity and temperature relaxation:

{{{%
\begin{equation}
I_{a}(\mathbf{p})=-\sum_{b}\nu _{ab}f_{a}^{0}(\mathbf{p})[\delta \mathbf{v}%
_{a}m_{a}\frac{\mathbf{V}_{a}\mathbf{-V}_{b}}{T_{a}}+\left( \frac{m_{a}}{%
T_{a}}\delta \mathbf{v}_{a}^{2}-3\right) \left( T_{a}-T_{b}\right) \frac{%
m_{a}}{m_{a}+m_{b}}].  \label{f.7}
\end{equation}%
The first term }}}in {(\ref{f.7}) describes relaxation of the momenta (}we
took into account the isothermal case) and the second term describes the
temperature relaxation (we neglected corrections due to the square mean
velocities). For this form of the collision integral {(\ref{f.7})} it is
easy to verify the Boltzmann H-theorem:

\begin{equation}
\frac{\partial }{\partial t}H(t)=\frac{\partial }{\partial t}\sum_{a}\int
f_{a}(\mathbf{p,}t)\ln f_{a}(\mathbf{p,}t)d\mathbf{p=}-\sum_{a}\int \frac{%
\delta \mathbf{p}^{2}}{2m_{a}T_{a}}I_{a}(\mathbf{p})d\mathbf{p=-}\frac{3}{4}%
\sum_{ab}\nu _{ab}\frac{m_{a}n_{a}}{m_{a}+m_{b}}\frac{\left(
T_{a}-T_{b}\right) ^{2}}{T_{a}T_{b}}\leq 0.  \label{f.10}
\end{equation}

Finally we can combine {(\ref{f.7}) with (\ref{eqn37}){{%
\begin{equation*}
I_{a}(\mathbf{p})=-\nu _{a}\left\{ f_{a}(\mathbf{p})-f_{a}^{0}(\mathbf{p})%
\left[ 1-\left( \frac{1-\Pr }{\Pr }\right) P_{aij}\frac{{\delta v}_{ai}{%
\delta v}_{aj}}{2P_{a}T_{a}}m_{a}\right] \right\}
\end{equation*}%
\begin{equation}
-\sum_{b}\nu _{ab}f_{a}^{0}(\mathbf{p})\delta \mathbf{v}_{a}m_{a}\frac{%
\mathbf{V}_{a}\mathbf{-V}_{b}}{T_{a}}-\sum_{b}\nu _{ab}f_{a}^{0}(\mathbf{p}%
)\left( \frac{m_{a}}{T_{a}}\delta \mathbf{v}_{a}^{2}-3\right) \left(
T_{a}-T_{b}\right) \frac{m_{a}}{m_{a}+m_{b}}.  \label{eqn50}
\end{equation}%
}}}

The time evolution of the many-component systems up to the hydrodynamic
stage can be described as follows: a first, local equilibrium components
achieves, and then a balance across the gas mixture comes. At all these
stages Boltzmann's H-theorem holds. Thus, the complicated exponential
dependence typical of the GK model appears to be unfounded and does not hold
for states remote from the full equilibrium.

\section{CONCLUSION}

Using the well-known projection technique, a new form of the collision
operator for the Boltzmann gas of hard spheres and for the Coulomb plasma
has been developed. The proposed collision operator takes into account
relaxation of the first 13 hydrodynamic moments properly and accounts for
the contribution of non-diagonal components in the expansion of the
linearized collision operator in the complete system of Hermite polynomials.
The non-diagonal components accounted for in this basis in the quadratic
approximation contribute to the diagonal components. It is shown that for a
system of charged particles with the Coulomb interaction potential, these
contributions are essential and comparable with Spitzer corrections to the
transport coefficients. In the case of the Boltzmann gas of hard spheres
these corrections are insignificant. In the case of a many-component system,
the nonlinear model collision integral is constructed on the basis of the
linearized one. Unlike previous cases, it does not exhibit any complicated
exponential dependence and avoids the coefficient ambiguity in the
many-component collision integral. Boltzmann's H-theorem is proved for our
model.

\section*{Acknowledgement}

A fruitful discussion with A. E. Shabad\ is gratefully acknowledged.


\end{document}